\documentclass[reprint,aps,prl]{revtex4-2}

\usepackage{graphicx,amssymb,amsmath}
\usepackage{color,multirow,array}
\usepackage[english]{babel} 

\newcommand{\be}{\begin{equation}}
\newcommand{\ee}{\end{equation}}
\newcommand{\bea}{\begin{eqnarray}}
\newcommand{\eea}{\end{eqnarray}}

\def\eq#1{Eq.~(\ref{#1})}

\begin{document}

\title{Quantum capacitance governs electrolyte conductivity in carbon nanotubes}

\author{Th\'eo Hennequin}
\author{Manoel Manghi}
\email{manoel.manghi@univ-tlse3.fr}
\affiliation{Laboratoire de Physique Th\'eorique, Universit\'e Paul Sabatier--Toulouse III, CNRS, France}
\author{Adrien Noury}
\author{François Henn}
\author{Vincent Jourdain}
\author{John Palmeri}
\email{john.palmeri@umontpellier.fr}
\affiliation{Laboratoire Charles Coulomb, Universit\'e de Montpellier, CNRS, France}

\date{\today}
\begin{abstract}
In recent experiments, unprecedentedly large values for the conductivity of electrolytes through carbon nanotubes (CNTs) have been measured, possibly owing to flow slip and a high pore surface charge density whose origin is still unknown. By accounting for the coupling between the \textit{quantum} CNT and the \textit{classical} electrolyte-filled pore capacitances, we study the case where a gate voltage is applied to the CNT. The computed surface charge and conductivity dependence on reservoir salt concentration and gate voltage are intimately connected to the CNT electronic density of states. This approach provides key insight into why metallic CNTs have larger conductivities than semi-conducting ones.

\end{abstract}

\maketitle

Although much experimental, theoretical, and molecular modeling effort has been devoted over the past years to understanding water and ion (electrolyte) transport through carbon nanotubes (CNTs)~\cite{Fornasiero_rev,Drndic_rev,Aluru_rev}, the origin of the electric charge localized on the surface of industrially important CNT based nanofluidic systems still remains unclear (see Ref.~\cite{JPCC} and references therein). It has already been proposed that this surface charge could arise from functional groups at the CNT entrances~\cite{SR,Nanoscale} and/or the specific adsorption of ions, such as OH$^{-}$~\cite{Grosjean}. Although the above cited studies lead to the conclusion that this surface charge plays a key role in governing ion transport in CNTs, it is difficult to regulate it directly and one is left to making inferences, for example by studying the variation of ionic conductance $G$ with the pH or salt concentration, $c_s$, of the external bulk reservoirs bounding the CNT. Intriguing results have been obtained, including a power law behavior, $G\propto c_s^\alpha$, with $1/2\leq\alpha\leq 1$, which could be interpreted as the manifestation of an underlying surface charge regulation mechanism~\cite{Secchi,Biesheuvel,PRE}.

Through a simplified feasibility study we propose in this work that by biasing a CNT incorporated in a nanofluidic system {\it via} an applied gate voltage, $V_g$, and taking into account explicitly the quantum capacitance (QC) of the quasi-1D CNT structure as well as the non-linear capacitance of the confined electrolyte ion the pore, it should be possible to quantify the CNT surface charge density $\sigma_{\rm Q}$ and establish a link between the intrinsic CNT electronic properties and ion transport through the same structure, such as the electrolyte conductance through the CNT. 
%Importantly, 
A major conclusion it that these intrinsic electronic properties will depend significantly, under certain conditions, on  whether the CNT is metallic (M) or semiconducting (SC).

The CNT quantum charge $\sigma_{\rm Q}$ arises from the low density of states (DOS) for the charge carriers (electrons and holes) in this quasi-1D system. Indeed, the applied $V_g$ creates a net charge on the CNT by perturbing the equilibrium occupation of the allowed energy levels.
Part of $V_g$ goes to raising the electric potential of the CNT to $\psi_0$ and part goes to shifting the chemical potential by $\Delta\mu= -eV_{\rm ch}=-e(V_{g} - \psi_0)$ (where $e$ is the absolute value of the electron charge): $\psi_0$ will then be lower than $V_g$ with (in the absence of other capacitances) the difference between the two giving rise to the chemical contribution, $V_{\rm ch}$.
The amplitude of the chemical potential shift is determined by $V_{g}$ and the relative values of the capacitances with the smallest one providing the most important contribution. 
We show here that for a wide range of salt concentration, $10^{-3}\leq c_s\leq 4$~mol/L, the CNT QC,  $C_{\rm Q}$, can be lower than the nanopore electrolyte contribution, $C_{\rm P}$, and therefore the quantum contribution controls the conductance.  
By comparing the confined electrolyte conductivity calculated for a classical metallic nanopore (with a very high,  effectively infinite, DOS)~\cite{Martin2014} with those calculated for M and SC CNTs, we bring to light clear signatures of each type of CNT behavior.

\begin{figure*}[t!]
\begin{center}
\includegraphics[width=1\textwidth]{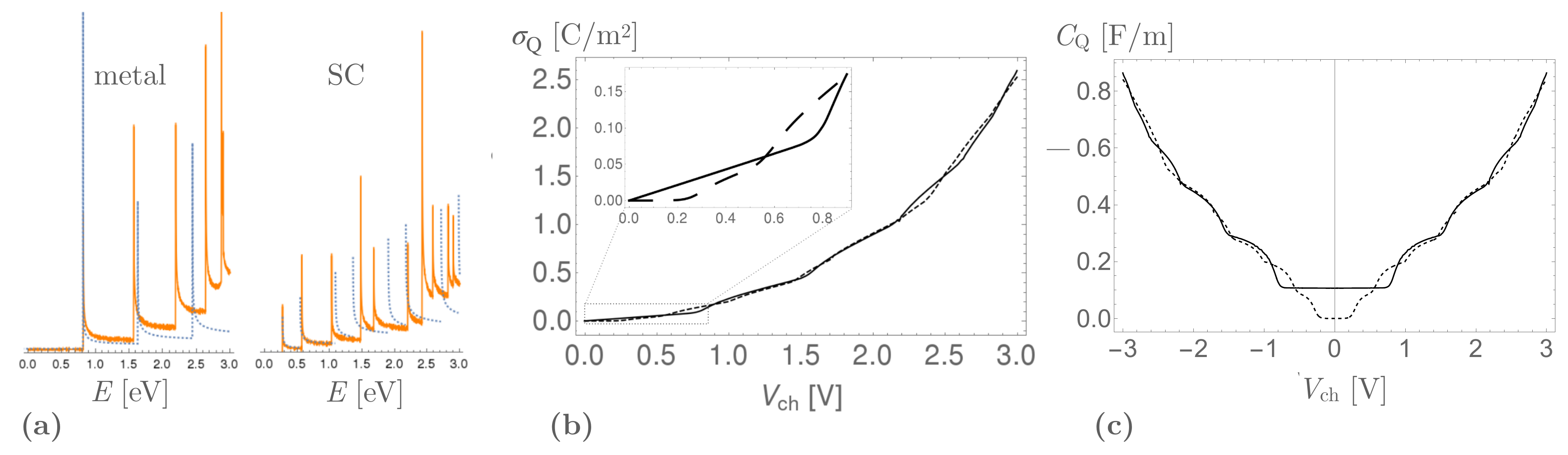}
\caption{(a)~Density of states (DOS) $g(E)$ for metallic (M,left) and  semi-conducting (SC,right) CNTs computed by the tight-binding method (solid orange lines) and the ${\bf k}\cdot{\bf p}$ approximation (blue dashed lines). (b)~Quantum surface charge density and (c)~integral QCs vs. the chemical contribution, $V_{\rm ch}$ for these M (solid curve) and SC (dotted curve) CNTs.
\label{fig1}}
\end{center}
\end{figure*}
The CNT QC has already been studied in setups where an electrolyte surrounds the CNT and acts as an electrolyte liquid gate, the main goal was however to investigate the electrical properties of the CNT itself~\cite{Rosenblatt2002,Heller2006,Heller2008,Heller2010,Burke2018,Burke2019}. Within this context, Lemay's group~\cite{Heller2006,Heller2008,Heller2010} has carried out an interesting study, although the capacitance of the external electrolyte was modeled using an overly simplified form (of the type used for Stern layers) that does not capture the full complexity of the non-linear capacitance of an electrolyte. The QC of graphene sheets and ribbons have also been studied theoretically and experimentally using both ionic liquid and electrolyte gates~\cite{Fang,Heller2010,Xia2009}.
The gating of a generic nanopore has also been studied theoretically for confined electrolytes in a nanofluidic setup where the gate voltage is applied to an electrode in contact with an insulating layer making up the finite-thickness surface of a nanopore~\cite{Jiang2010,Jiang2011}. The confined electrolyte conductance was investigated, but the important special case of a CNT was not addressed: the system was treated as being purely classical and therefore the capacitances considered were the geometrical one coming from the insulating layer, together with the Stern and  non-linear Graham capacitances of the electrolyte.

In contrast to these previous studies we show here that by examining the confined electrolyte conductivity the CNT QC can provide a direct adjustable handle on the all important, but elusive, nanopore surface charge density. By shedding light in a fundamental way on ion transport in this type of nanofluidic system, it should be possible to pave the way to the tuning of transport properties for particular practical applications.

We briefly summarize the electronic characteristics of single-walled carbon nanotubes (SWCNTs) obtained by simplified theoretical methods and tight-binding calculations (see Ref.~\cite{McEuen} for more details). A SWCNT can be considered as a graphene sheet rolled up into a cylinder and its electronic nature, metallic (M) or semi-conducting (SC), depends on its atomic structural characteristics fixed by two \emph{chiral} indices $(n,m)$ that determine its diameter and chirality~\cite{Mintmire,Heller2006,Ilani,McEuen}.
The simplified theoretical ${\bf k}\cdot{\bf p}$ method, which neglects curvature effects, reveals that in the M case, $|n-m|=3q$, where $q\in \mathbb{N}$. In the SC case, $|n-m|\neq 3q$ and the size of the band gap is inversely proportional to the SWCNT diameter $d$~\footnote{the inclusion of curvature in more sophisticated methods shows that only armchair SWCNTs are truly metallic, the other putatively metallic SWCNTs pick up a small band gap that can be considered for our purposes to be smeared out at room temperature}.
The band structure is composed of multiple 1D subbands sliced from the Dirac dispersion cone of graphene and, as shown by simplified theoretical analyses~\cite{Mintmire}, exhibits a certain amount of (approximate) universality. The $i$th electron-hole subband has the following dispersion relation
\be
E_\pm(i,k)= \pm\sqrt{(\hbar v_F k)^2+E_i^2},
\ee
where $v_F\approx 8\times 10^5~$m/s is the Fermi velocity at the Dirac cone, $\hbar$ the reduced Planck's constant, $k$ the charge carrier wave vector, and $E_i=2 \hbar v_F i/(3d)$. The allowed values of $i$ depend on whether the SWCNT is M or SC~\cite{Mintmire,Miyake}.
The $E_{i,g}=2 E_i$ are the individual energy gaps determined by the distance of the quantized $k$ states from the center of the Dirac cone situated where both $k$ and $E$ vanish.
The DOS per unit surface (the usual DOS per unit length divided by $\pi d$), for a SC SWCNT is given by~\cite{Mintmire,Heller2006,Ilani,McEuen}
\be
g_{\rm SC} (E) = g_0 \sum_{i>0, \neq 3 q} \frac{|E|}{\sqrt{E^2-E_i^2}} \Theta(|E|-E_i),
\ee
where $g_0 =4/(\pi^2 d \hbar v_F)$ and $\Theta$ is the Heaviside unit step function. The SC band gap is predicted to be $E_g=E_+(1,0)-E_-(1,0)=2E_1=0.7/d$~eV, if $d$ is in nm (between 0.2 to 0.7~eV for $1\leq d\leq3$~nm). For purposes of illustration we choose here $d=1.5$~nm for which $E_g=0.47$~eV (18~$k_B T$ at room temperature).
The subbands for  $i=3 q$  (integer $q$) are not present [see Fig.~\ref{fig1}(a)] and therefore the SC DOS exhibits van Hove singularities at $|E|= E_i$ for $i >0, i\neq 3 q$.

For true M SWCNTs,  $E_g=0$ and the first band has a non-zero DOS $g(0)=g_0$ and a half-width equal to $E_3$ as shown in Fig.~\ref{fig1}(a)]~\cite{McEuen}. 
The ${\bf k}\cdot{\bf p}$ method leads to a DOS given by~\cite{Mintmire}
\be
g_{\rm M} (E) =  g_0 + 2 g_0 \sum_{m>0} \frac{|E|}{\sqrt{E^2-(m E_3)^2}} \Theta(|E|-m E_3).
\ee
The M DOS exhibits van Hove singularities at $|E| = m E_3$ for $m >0$ and, aside from the conduction band, all bands are doubly degenerate.

We show in Fig.~\ref{fig1}(a) the approximate theoretical DOS for $d=1.5$~nm (dashed blue curves) and the DOS computed by the group of Maruyama~\cite{Maruyama} using the tight-binding method (solid orange curves) for a M CNT of chiral indices $(11,11)$ ($d=1.512$~nm) and for a SC CNT of chiral indices $(19;0)$ ($d=1.508$~nm). It is clear that the approximate DOS universality breaks down for $|E|>1$~eV where the actual DOS begins to depend on the values of the chiral indices.
 
At thermal equilibrium without external perturbations the chemical potential is situated at the Dirac point taken as the reference $\mu = 0$~eV. When a gate voltage $V_g$ is applied to an electrode posed on a CNT, a shift in the chemical potential, $\Delta\mu$, occurs and the occupation of the allowed levels is modified, which leads to the creation of charges (electrons or holes) in the various subbands labeled by $i$. The CNT DOS plays 
%an important 
a crucial role in determining the amplitude of this charge. The density of electrons (number per unit surface), $n$, for a SWCNT is calculated as a function of temperature $T$ and $\Delta\mu$ using Fermi statistics~\cite{Mintmire,Heller2006,Ilani,McEuen},
\be
n(\Delta\mu) = \int_{E_g/2}^\infty \frac{g(E)}{1+\exp[\beta (E-\Delta\mu)]} dE,
%p(\Delta\mu) &=& \int_{E_g/2}^\infty \frac{g(E)}{1+\exp[\beta (E+\Delta\mu)]} dE,
\label{n}
\ee
where $\beta = 1/(k_B T)$ 
($k_B T \approx 25$~meV at room temperature). 
The density of holes, $p$, is given by \eq{n} by replacing $-\Delta\mu$ by $\Delta\mu$.
The \textit{chemical contribution} to the applied gate voltage (electromotive force), $V_{\rm ch} = V_g-\psi_0$
engenders a charge per unit surface, 
%$\sigma_{\rm Q}$, 
\be
\sigma_{\rm Q}(V_{\rm ch}) = e(p-n),
\label{sigmaQ}
\ee
and is related to the chemical potential shift through $\Delta\mu=-eV_{\rm ch}$, where $\psi_0$ is the electrostatic potential drop between the CNT and the reference electrode. 
Note that $\sigma_{\rm Q}=0$ for $V_{\rm ch}=0$, $\sigma_{\rm Q}>0$ for $V_{\rm ch}>0$, and $\sigma_{\rm Q}<0$ for $V_{\rm ch}<0$~\cite{Zhang,Berthod}.
The surface charge densities $\sigma_{\rm Q}(V_{\rm ch})$ for SC and M CNTs are plotted in Fig.~\ref{fig1}(b). For $V_{\rm ch}>1$~V, they almost superimpose, showing some slope discontinuities associated with the van Hove singularities of the associated DOS.  
As seen in the inset, however, for $V_{\rm ch}<1$~V the two are quite different, essentially because the charge density vanishes in the gap for the SC CNT, i.e. for $V_{\rm ch}<E_g/(2e)=0.23$~V.

We distinguish between the \textit{integral} QC needed here, 
\be
C_{\rm Q}\equiv\frac{|\sigma_{\rm Q}|}{V_{\rm ch}},
\label{CQI}
\ee
which directly determines the charge 
%$\sigma_{\rm Q}$ 
on the CNT,
and the \textit{differential} one, $C^d_Q(V_{\rm ch}) \equiv \partial |\sigma_{\rm Q}|/\partial V_{\rm ch}$,
which gives access directly to many features of the DOS.
The integral capacitances for SC and M CNTs are plotted in Fig.~\ref{fig1}(c). 
If $|V_{\rm ch}|<E_3/e$ holds, then $C_{\rm Q}\simeq C^0_Q$ for M CNTs and $C_{\rm Q}=0$ for SC ones. % [using $S(0,\eta)-S(0,-\eta)\approx\eta$]. \\
%As for 
For larger $|V_{\rm ch}|$, similarly to $\sigma(V_{\rm ch})$, the two integral capacitances are quite similar.

\begin{figure*}[t]
\begin{center}
\includegraphics[width=1\textwidth]{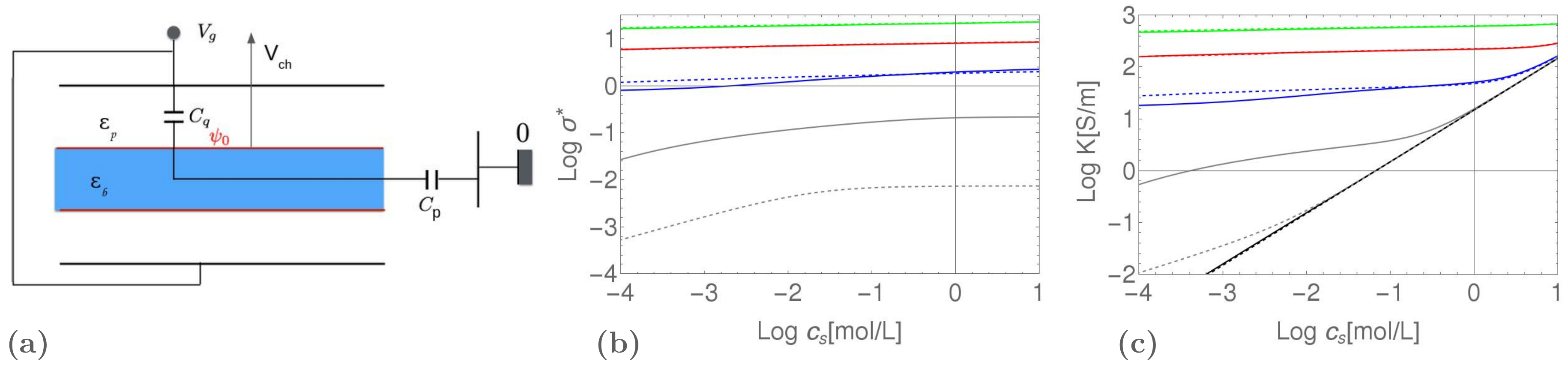}
\caption{(a)~Equivalent electrical circuit of the system with the quantum capacitance $C_{\rm Q}$ and the pore one $C_{\rm P}$. The gate voltage, $V_g =\psi_0+V_{\rm ch}$, is divided between the potential at the inner CNT surface, $\psi_0$, and the chemical contribution, 
$V_{\rm ch}$.
%$V_{\rm ch}=V_g-\psi_0$. 
(b)~Dimensionless surface charge density $\sigma^*$ and (c)~conductivity of a KCl electrolyte vs. salt concentration $c_s$ (Log-Log plot) for the M (solid lines) and SC (dashed lines) CNTs with $V_g=0.2$ (grey), 1 (blue), 2 (red), and 3~V (green). The thick line is the bulk conductivity. 
\label{fig2}}
\end{center}
\end{figure*}

In the simplified description we adopt here (absence of any residual geometrical capacitance arising from the rest of the electrical circuit) the gate voltage, $V_g = V_{\rm ch} + \psi_0$, is shared between the chemical contribution and the electrostatic one. The equivalent electrical circuit is shown in Fig.~\ref{fig2}(a), the total capacitance, $C_{\rm tot}$, and amplitudes of each contribution to $V_g$ depend on the individual capacitances and are given by~\cite{Heller2006,Fang,Zhang,Berthod}:
\be
\frac{1}{C_{\rm tot}} = \frac{1}{C_{\rm Q}} + \frac{1}{C_{\rm P}} = \frac{V_g}{|\sigma_{\rm Q}|},
\ee
where $C_{\rm P}=|\sigma_{\rm Q}|/\psi_0$ is the non-linear integral \textit{pore} capacitance arising from  electrolyte confined to the interior of a CNT,
which leads to:
\be
\psi_0 = V_g\frac{C_{\rm Q}(V_{\rm ch})}{C_{\rm Q}(V_{\rm ch})+C_{\rm P}(\psi_0)}.
\label{defpsi0}
\ee
%\eq{defpsi0} 
This result shows clearly that if $C_{\rm Q}\gg C_{\rm P}$ then $\psi_0\lesssim V_g$ and $V_{\rm ch}\ll V_g$, 
indicating that
quantum effects are negligible. It corresponds to the ``classical'' limit. Conversely, if $C_{\rm P}\gg C_{\rm Q}$, then $\psi_0\ll V_g$ and quantum effects are dominant.

To calculate the electrolyte capacitance, $C_{\rm P}$, we use an interpolation formula for the electric potential at the pore wall, $\psi_0$, that we previously devised for small nanopores~\cite{PRE}. This formula for $\psi_0$ as a function of the nanopore radius $R$ ($=d/2$),  electrolyte salt concentration $c_s$,  and nanopore surface charge density, $\sigma_{\rm P}$, becomes exact in the good coion exclusion  and homogeneous limits and is an accurate interpolation elsewhere:
\be
\sinh\tilde \psi_0 = \frac{\sigma_{\rm P}^*(1+\sigma_{\rm P}^*)}{\tilde c_s},
\label{c}
\ee
where 
$\tilde \psi_0=\beta e |\psi_0|$
and
\be
\tilde c_s= \pi \ell_B R^2 c_s, \qquad\sigma_{\rm P}^*=\frac{\pi R\ell_B}{e}|\sigma_{\rm P}|
\ee
are respectively the dimensionless electrolyte salt concentration and
surface charge density
(the electric potential at the reference electrode in the bulk reservoir is taken to be zero). 
The Bjerrum length is $\ell_B=e^2/(4\pi\epsilon k_BT)\simeq 0.7$~nm in bulk water at room temperature (dielectric permittivity $\epsilon=78\epsilon_0$). \eq{c} can be rewritten by isolating $\sigma_{\rm P}^*$ as
\be
\sigma_{\rm P}^* = \frac12\left(\sqrt{1+4\tilde c_s\sinh\tilde \psi_0}-1\right) \label{sigmaP}.
\ee
For a classical metal (effectively infinite DOS), formally $C_{\rm Q}/C_{\rm P}\to\infty$ and using \eq{defpsi0} we see that the surface charge density $\sigma_{\rm cl}^\mathrm{*}$ is given by \eq{sigmaP} with $\psi_0$ set to $V_g$. In this case for low enough $\tilde c_s$ and $V_g$,
$\sigma_{\rm cl}^* \simeq \tilde c_s\sinh(\beta e V_g)$ and for high enough $\tilde c_s$ and $V_g \gg 25$~mV, 
$\sigma_{\rm cl}^* \simeq \tilde c_s^{1/2}e^{\beta e V_g/2}$.
The rapid exponential rise in surface charge density in the latter limit appears to be compatible with the extremely high conductivities observed in \cite{Martin2014} for gold plated nanopores at negative gate voltages (for which specific ion adsorption can be neglected). Given the low DOS of CNTs we do not expect such classical exponential behavior for them. In this case
by keeping $\sigma_{\rm P}^*$ constant, $\tilde \psi_0$ is low when $\tilde c_s$ is high, which increases the pore capacitance and eventually leads to $C_{\rm P}\gg C_{\rm Q}$, i.e. quantum effects become important.
%For $\tilde \psi_0 \gg 1$ and

By defining $C_{\rm P}=\frac{4\epsilon}{R} \tilde C_{\rm P}$, the integral pore capacitance can be found from the dimensionless one:
\be
\tilde C_{\rm P} =\frac{\sigma_{\rm P}^*}{\tilde\psi_0}=\frac{\sqrt{1+4\tilde c_s\sinh\tilde \psi_0}-1}{2\tilde\psi_0}
\ee
For $\tilde\psi_0 \ll 1$, $\tilde C_{\rm P} \rightarrow \tilde c_s$ and $ C_{\rm P} \rightarrow \epsilon R/(2 \lambda_{\rm DH}^2)$, where $\lambda_{\rm DH} =(8\pi\ell_B c_s)^{-1/2}$ is the bulk Debye-H\"uckel screening length for a monovalent electrolyte. In this limit the electrolyte filled pore acts as a parallel plate capacitor of capacitance $\epsilon/d_{\rm eff}$ with effective thickness $d_{\rm eff} = 2 \lambda_{\rm DH}^2/R$.

Now by writing $\sigma\equiv\sigma_{\rm Q}=\sigma_{\rm P}$ and combining $\sigma_{\rm Q}(V_g-\psi_0)$ from \eq{sigmaQ} and \eq{c}, we obtain $\tilde c_s(V_g,\psi_0)$, which enables us to trace a parametric plot of $\sigma$ as a function of $c_s$ for fixed gate voltage $V_g$ by varying $\psi_0$ from 0 to $V_g$. 
This is shown in Fig.~\ref{fig2}(b) for the SC and M cases  for $V_g=0.2,1,2$ and 3~V. For $V_g=0.2$~V, the surface charge density is two orders of magnitude larger for the M CNT than for the SC one. This is because $V_g$ is just below $E_g/2$ (gap half-width). 
At this relatively low value of $V_g$ the surface charge density increases quickly with $c_s$ at low concentrations. For higher gate voltages, $\sigma$ increases very slowly with $c_s$, being almost constant for $V_g \geq 2$~V. 
For $V_g=1$~V (blue curves) $\sigma({\rm SC})>\sigma({\rm M})$ at low $c_s$ because $\psi_0$ is high and $V_{\rm ch}$ is too low  to reach beyond the SC van Hove singularity occurring at $E_g/(2e)=0.23$~V [see Fig.~\ref{fig1}(a)]. For large $c_s$, however, $\psi_0$ becomes lower and $V_{\rm ch}\simeq E_3/e=0.7$~V  higher, now reaching the large first van Hove singularity for the M case.

The pore conductance $G$ is related to the conductivity $\kappa$ through $G=\pi R^2\kappa/L$ where $L$ is the pore length. The pore conductivity, $\kappa$, as a function of $\sigma^*$ and $\tilde c_s$ can be obtained using the result obtained in Eq.~19 of Ref.~\cite{PRE}, which takes into account the electrical migration contribution (first two terms) and the electro-osmotic one (third term), including flow slip at the pore surface:
\begin{multline}
    \tilde \kappa =(\tilde\mu_+ +\tilde\mu_-)\tilde c_s \sqrt{1+\left(\frac{\sigma^*}{\tilde c_s}\right)^2} 
    -\mathrm{sgn}(\sigma)\sigma^* (\tilde\mu_+ -\tilde\mu_-)\\
    + 2 \sigma^* \left[1-\frac{\ln(1+\sigma^*)}{\sigma^*}+2\tilde b \sigma^*\right],
\label{kappa}
\end{multline}
where $\tilde \kappa =(2\pi^2 R^2\ell_B^2 \eta\kappa/e^2) $ is the dimensionless conductivity ($\eta=8.94\times10^{-4}$~Pa.s being the water viscosity), $\tilde\mu_\pm=2\pi\eta\ell_B\mu_\pm$ the dimensionless mobilities of ions (taken equal to the bulk values, $\mu_+\approx\mu_-=5\times 10^{-11}$~s/kg for KCl), and $\tilde b= b/R$ is the ratio between the slip length $b$ and the pore radius [$\mathrm{sgn}(\sigma)$ denotes the sign of $\sigma$]. In \eq{kappa} the conductivity $\kappa(c_s,V_g)$ is a function of both $c_s$ and $V_g$ because $\sigma(c_s,V_g)$ is a function of both these quantities.
\begin{figure*}[t!]
\begin{center}
\includegraphics[width=1\textwidth]{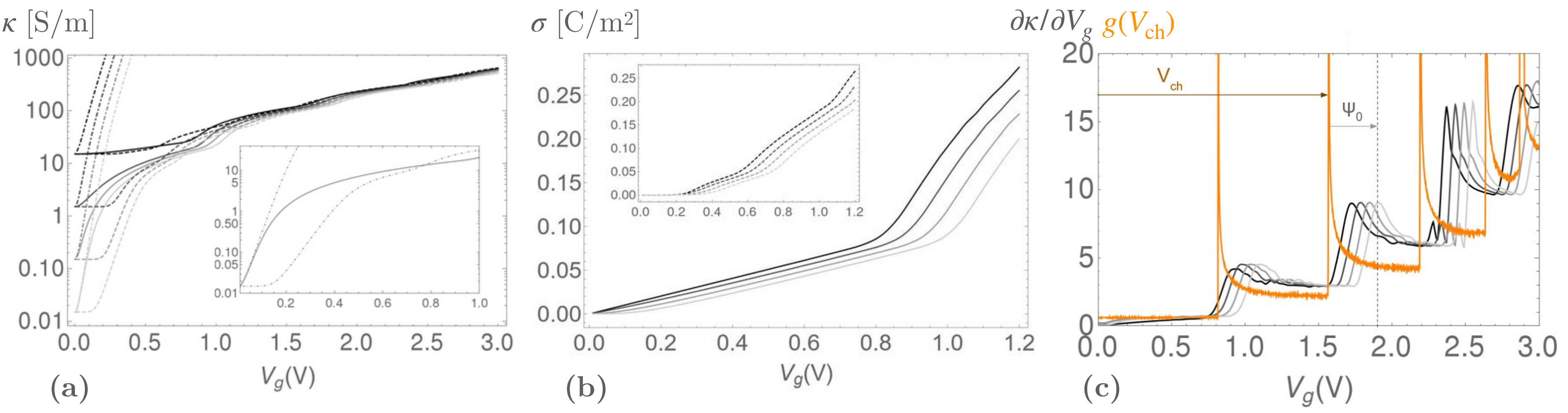}
\caption{
(a)~Electrolyte conductivity $\kappa$ vs. gate voltage $V_g$ in M (solid lines) and SC (dashed lines) CNTs for reservoir salt concentration values of (from bottom to top) $c_s = 10^{-3},10^{-2},10^{-1},1$~mol/L. The classical metallic CNT is shown in dotted-dashed lines. The inset is a zoom for $c_s = 10^{-3}$~mol/L. (b)~Associated surface charge density vs. $V_g$ for the M CNT for the same $c_s$ values as in~(a). The SC CNT is shown in the inset. 
(c)~Derivative of the electrolyte conductivity, $\kappa'(V_g)$, in the M CNT for the same $c_s$ values as in~(a). The DOS $g(V_{\rm ch})$ is superimposed in orange. To illustrate how to read off $\psi_0$ we show 
it and $V_{\rm ch}$ for the case $c_s=10^{-3}$~mol/L and $V_g=1.9$~V.}
\label{fig3}
\end{center}
\end{figure*}

One observes in Fig.~\ref{fig2}(c), that the conductivity is enhanced compared to the bulk case for intermediate and low salt concentrations. This is the signature of a non-zero surface charge density.
%and an enhancement of the counter-ion concentration in the nanopore. 
%For $V_g>1$~V it is even true for large $c_s$. 
For $V_g>1$~V, this enhancement even persists  to high  $c_s$, attesting to a high surface charge density.
At low salt concentration, $\kappa$ in \eq{kappa} is controlled by $\sigma$ (good coion exclusion regime) and we  therefore  observe
in Fig.~\ref{fig2}(c)
an increase in conductivity similar to the one observed for $\sigma(c_s)$. 
We have checked that for $V_g>1$~V, $\sigma^*>\tilde c_s$ over the whole 
%any $c_s$ 
concentration range and $\kappa$ is therefore controlled by the first electrical migration term in \eq{kappa}, which is one order of magnitude larger than the electro-osmotic one (the second term is negligible since $\mu_+\approx\mu_-$ for KCl). For the M case and $V_g<1$~V (solid grey line), $\kappa$ increases with $c_s$, but much more slowly than the bulk one, with an apparent exponent smaller than $1/2$. 

By taking a slip length large enough, the electro-osmotic contribution in  \eq{kappa} associated with the fluid slip at the nanopore surface,  $4\tilde  b\sigma^{*2}$, can eventually dominate the electrical migration one. For instance, if a reasonable value of $\tilde b=40$~\cite{Kannam2017} is taken, the electro-osmotic contribution is greater by one order of magnitude over the whole concentration range.
The conductivity would then be completely controlled by $\sigma^*(c_s)$ [plotted in Fig.~\ref{fig2}(b)], as a more complete study (underway) will reveal.

Changing the concentration in micrometric reservoirs is not easy in practice. 
One could measure more easily the electrolyte conductivity at fixed $c_s$ by varying $V_g$. 
%To do so, 
To present the conductivity in this way, we first compute $\psi_0$ by solving numerically $\sigma_{\rm Q}(V_g-\psi_0)=\sigma_{\rm P}(c_s,\psi_0)$ using Eqs.~(\ref{sigmaQ}) and (\ref{sigmaP}) at fixed $V_g$ and $c_s$. We then re-inject $\psi_0$ in $\sigma_{\rm P}$, which itself can finally be injected into \eq{kappa}.
In Fig.~\ref{fig3}(a) we plot the conductivity as a function of 
$V_g$
%$\kappa(V_g)$ 
at fixed $c_s=10^{-3},10^{-2},10^{-1},1$~mol/L from bottom to top. We confirm that $\kappa$ varies strongly at low gate voltages.  For $V_g=0$, $\sigma=0$ and $\kappa\propto c_s$ (bulk conductivity), which explains the staggered values (separated by a decade) at low $V_g$.   The distinction between the three cases, semiconductor, quantum metal and ``classical'' metal is clear at low $V_g$, where the QC suppresses the ``classical'' exponential growth of conductivity with $V_g$ for M and SC CNTs and the SC energy gap suppresses the conductivity with  respect to that of the quantum metal. The M CNT conductivity follows the classical one at very low $V_g$ within a window the decreases in width with increasing $c_s$. 
At high $V_g$ all M and SC curves are almost superimposed.

Even if the quantum surface charge density $\sigma$, has the same derivative as the classical one at $V_g=0$, it becomes quickly linear 
$\sigma \simeq C_{\rm Q}^0 V_g$, with a lower slope up to $V_g\simeq E_3/e\simeq 0.7$~V, as shown in Figure~\ref{fig3}(b). 
Indeed, for not too low $V_g$, $C_{\rm Q}\ll C_{\rm P}$, which implies that $C_{\rm tot}\simeq C_{\rm Q}\approx C_{\rm Q}^0$ and therefore $\psi_0\ll V_g$. We have checked that this is true whenever $C_{\rm P}/C_{\rm Q}^0\gg 1$, which leads to $\sqrt{\tilde c_s}>\tilde C_{\rm Q}^0$. Since $\tilde C_{\rm Q}^0=0.03$, this last inequality is verified for $\tilde c_s>10^{-3}$ [the case for the concentrations shown in Fig.~\ref{fig3}(a)]. For lower reservoir salt concentrations, $\sigma$ is no longer linear in $V_g$.
These different behaviors are illustrated in the inset of Fig.~\ref{fig3}(a) where we also note the impact of the SC CNT gap, which leads to a constant $\kappa$ up to $V_g=0.2\simeq E_g/e$~V, the edge of the gap. 
We conclude that the QC has a considerable effect on both $\sigma$ and $\kappa$ as soon as $V_g>k_BT/e\approx 25$~mV and this  over the whole experimental range of $c_s$. 

The structure of  a conductivity curve $\kappa(V_g)$ exhibiting slope changes, related to the DOS structure, is easier to visualize by plotting its derivative $\kappa'(V_g)$ [see Fig.~\ref{fig3}(c)] and superimposing on it the DOS $g(V_{\rm ch})$ (in orange). Clearly the peaks in $\kappa'(V_g)$ are directly connected to the van Hove singularities of the DOS with an offset to the right which increases when $c_s$ decreases. Indeed, since $\kappa$ is governed by the surface charge density $\sigma_{\rm Q}$ which depends only on $V_{\rm ch}=V_g-\psi_0$, the higher $\psi_0$ is (and therefore the lower $c_s$ is), the more $\kappa(V_g-\psi_0)$ is shifted to the right. We can therefore directly measure the potential $\psi_0$ in Fig.~\ref{fig3}(c). %as the abscissa of the conductivity pic minus the one of the nearest to its left van Hove singularity. 
For instance, for $c_s=10^{-3}$~mol/L and $V_g=1.9$~V, we get $\psi_0\simeq 0.32$~V. This is a way to readily check that for these $c_s$ values, $\psi_0\ll V_g$, thus confirming the central role played by the QC on the conductivity.

%CONCLUSION

As an concluding example, we consider the experimental results obtained by Liu et al.~\cite{Liu2013}, who measured, for $c_s=1$~mol/L, conductances of 61.0~nS for M SWCNTs and 5.6~nS for SC ones with $0.8\leq d\leq2$~nm and $5\leq L\leq10$~nm. Using the DOS given in Fig.~\ref{fig1}(a) with $d=1.5$~nm and $L=8$~nm, we can account for these two conductance values by taking $V_g\simeq 0.35$~V, 
a value that we interpret as an environmentally induced shift in the zero of the gate tension.
%An interpretation would be that the electrostatic environment of the experimental setup would have fixed the CNT at a gate tension of 0.35~V.

More systematic experiments are clearly needed to ascertain to what extent the charge density and therefore the electrolyte conductivity through SWCNTs can be controlled by an applied gate voltage. 
Presumably, a more  complete model for conductivity would be needed to account for the full complexity of real CNTs, including the influence of pH (via charge regulation), residual geometrical capacitances, and dielectric interactions.
%Presumably, a more for a complete model for conductivity including the influence of pH, one might add a part of $\sigma$ coming from a charge regulation.

\end{document}